\documentclass{article}
\usepackage{spconf,amsmath,graphicx}

\usepackage[colorlinks=true, linkcolor=blue, citecolor=red, urlcolor=magenta]{hyperref}

\usepackage{cite}
\usepackage{amssymb,amsfonts}
\usepackage{algorithmic}

\usepackage{textcomp}
\usepackage{xcolor}
\usepackage{booktabs}
\usepackage{makecell}
\usepackage{threeparttable}
\usepackage{subcaption}
\usepackage{colortbl,xcolor}

\usepackage[capitalize]{cleveref}
\crefname{section}{Sec.}{Secs.}
\Crefname{section}{Section}{Sections}
\Crefname{table}{Table}{Tables}
\crefname{table}{Tab.}{Tabs.}

\usepackage{multirow}
\usepackage{rotating}

\title{UnlearnShield: Shielding Forgotten Privacy against Unlearning Inversion}
\name{ \shortstack[l]{
    Lulu Xue\textsuperscript{1}, Shengshan Hu\textsuperscript{1}, Wei Lu\textsuperscript{1}\thanks{Minghui’s work is supported in part by the National Natural Science Foundation of China (Grant No.62572206).  * is the corresponding author.},
    Ziqi Zhou\textsuperscript{1}, Yufei Song\textsuperscript{1}*, \\
    Jianhong Cheng\textsuperscript{2},  Minghui Li\textsuperscript{1}, Yanjun 
  Zhang\textsuperscript{3}, Leo Yu Zhang \textsuperscript{4}
  }}

\address{$^{1}$ 
Huazhong University of Science and Technology,\\
$^{2}$ Institute of Guizhou Aerospace Measuring and Testing Technology,\\
$^{3}$ University of Technology Sydney, $^{4}$ Griffith University.\\
{\footnotesize\texttt{\{lluxue,hushengshan,luwei\_hustcse,zhouziqi,yufei17,minghuili\}@hust.edu.cn}}\\
{\footnotesize\texttt{jianhong\_cheng@csu.edu.cn},}{\footnotesize\texttt{Yanjun.Zhang@uts.edu.au},}{\footnotesize\texttt{leo.zhang@griffith.edu.au}}
}

\begin{document}
\maketitle
\begin{abstract}
Machine unlearning is an emerging technique that aims to remove the influence of specific data from trained models, thereby enhancing privacy protection. However, recent research has uncovered critical privacy vulnerabilities, showing that adversaries can exploit unlearning inversion to reconstruct data that was intended to be erased. Despite the severity of this threat, dedicated defenses remain lacking. To address this gap, we propose UnlearnShield, the first defense specifically tailored to counter unlearning inversion. UnlearnShield introduces directional perturbations in the cosine representation space and regulates them through a constraint module to jointly preserve model accuracy and forgetting efficacy, thereby reducing inversion risk while maintaining utility. Experiments demonstrate that it achieves a good trade-off among privacy protection, accuracy, and forgetting.
\end{abstract}

\section{Introduction}
\label{sec:intro}
Machine unlearning~\cite{GA1} is an emerging technique that removes the influence of specific data from trained models to enhance user privacy. It typically involves two models: the original model and the unlearned model produced by an unlearning algorithm. Prior studies~\cite{if,GA1} have shown that the resulting parameter difference between the two models ({DiffParm}) may retain gradient information related to the forgotten data. This observation has motivated the exploration of potential attack vectors that exploit {DiffParm}.
\begin{figure}[t]
 \setlength{\abovecaptionskip}{4pt}
    \centering
    \includegraphics[scale=0.27]{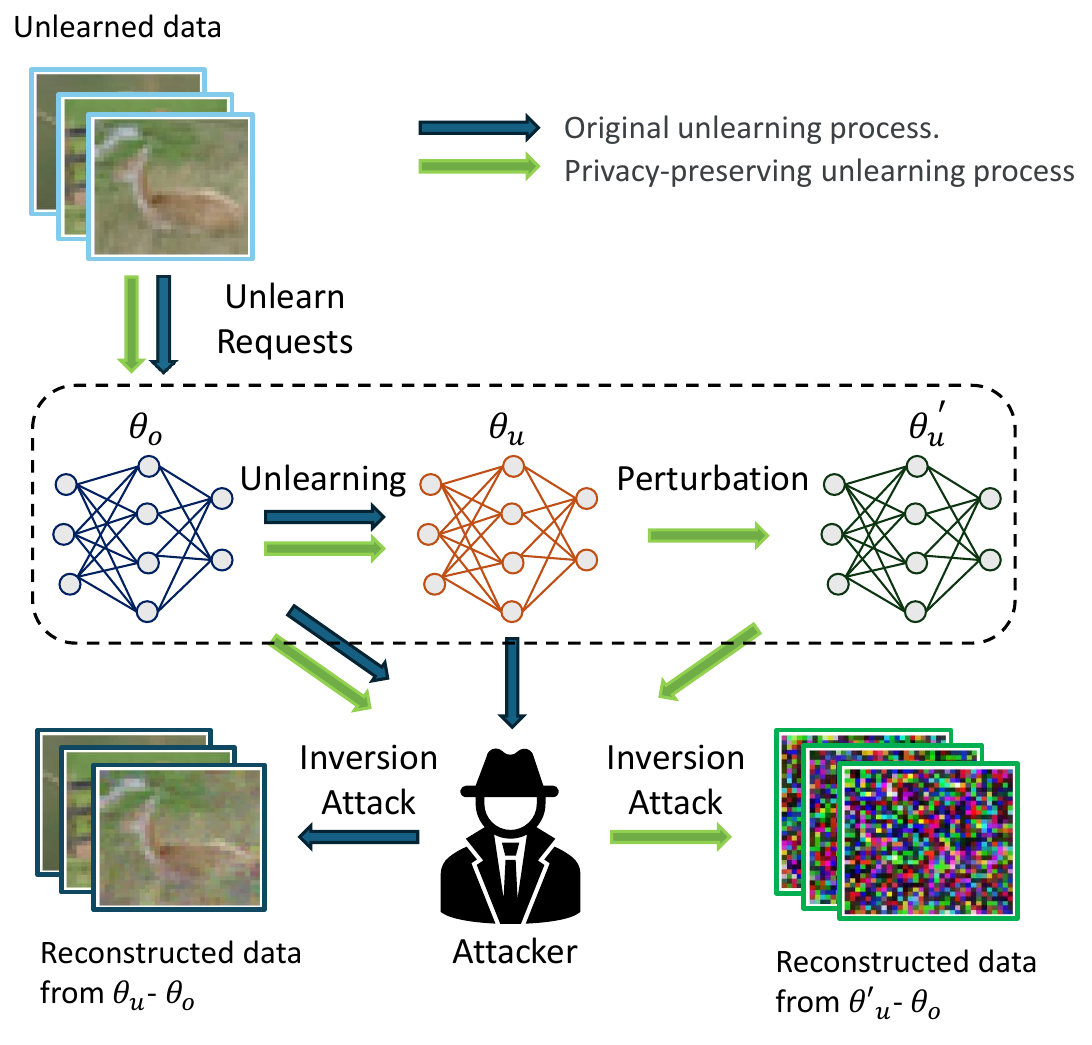}
    \caption{
    Post-processing defense pipeline against UIA.
    }
    \label{fig:demo}
  \vspace{-1.6em}
\end{figure}
Building on this insight, recent work~\cite{recon1} introduces the concept of {unlearning inversion attack (UIA)}, showing that attackers can exploit these gradient signals in DiffParm to reconstruct data that should have been erased, thereby compromising the privacy guarantees of unlearning. However, no dedicated defenses have been developed to address this threat. To fill this gap, we apply several gradient privacy protection methods~\cite{soteria,outpost,dgp,sok} to UIA. However, we observe that these defenses fail to prevent effective reconstruction. This is because DiffParm often encodes sensitive information about the forgotten data in the cosine space, whereas existing methods lack targeted designs for this space and primarily operate by perturbing gradient magnitudes in the euclidean space.

To address this challenge, we propose UnlearnShield, a post-processing defense tailored for UIA, as illustrated in Fig.~\ref{fig:demo}. UnlearnShield aims to achieve two goals: \textbf{privacy protection} and \textbf{utility preservation}.
For privacy protection, UnlearnShield injects a perturbation into the cosine space of DiffParm to enlarge the angular distance between the original and perturbed DiffParm, thereby disrupting reconstruction-critical directions.
For utility preservation, we consider two aspects: the accuracy of the unlearned model and its forgetting effect. To maintain model accuracy, UnlearnShield incorporates a magnitude constraint module in its perturbation optimization. To preserve the forgetting effect, we draw on the observations in~\cite{Salun} and introduce a forgetting-preserving function to ensure that the model retains its original forgetting behavior after perturbation.
In summary, our contributions are as follows:
1) We present the first exploration of defenses against unlearning inversion, enhancing privacy in unlearning scenarios.
2) We propose UnlearnShield, a post-processing defense that mitigates such attacks while preserving accuracy and forgetting performance.
3) Experiments demonstrate our method’s effectiveness across diverse datasets and models.

\section{Related Works}
\label{related_work}
\textbf{{Unlearning inversion attack.}} 
Machine unlearning~\cite{mu_new,mu_new1} enables models to forget the influence of specific data, supporting users’ right to be forgotten. To improve usability, various approximate methods have been proposed, with gradient-based techniques~\cite{GA1,sparse,Salun} proving particularly effective. However, recent work~\cite{recon1} reveals that such methods face severe privacy risks through the proposed Unlearning inversion attack (UIA). Given the original trained model $\mathbf{\theta}_o$, the unlearned model $\mathbf{\theta}_u$, and the forgotten data $\mathbf{x}$, We define the parameter difference, {DiffParm}, as $\Delta = \mathbf{\theta}_u - \mathbf{\theta}_o$.
The attacker aims to find a virtual input $\mathbf{x}'$ that minimizes the distance between the original and simulated parameter differences:
\[
\arg\min_{\mathbf{x'}} D(\Delta, \Delta^1),
\]
where \(D\) is a distance function, \(\mathbf{\theta}_u^1\) denotes the unlearned model generated by \(\mathbf{x}'\), and \(\Delta^1 = \mathbf{\theta}_u^1 - \mathbf{\theta}_o\).
As analyzed in~\cite{recon1}, DiffParm leaks privacy as it encodes gradient information of $\mathbf{x}$. However, since these gradients are nearly zero in the $\mathbf{\theta}_o$, euclidean optimization becomes ineffective~\cite{recon1,IG}. Therefore, attackers adopt cosine distance as $D$ to reconstruct the forgotten data.


\noindent\textbf{{Gradient privacy protection}}
Existing privacy protection strategies for gradient information can be broadly categorized into two types: {data-mixing} and {perturbation}-based. Data-mixing approaches~\cite{evaluating,concealing} aggregate multiple samples during training and are thus incompatible with fine-grained unlearning (e.g., deleting a single image). In contrast, perturbation-based defenses preserve data granularity by modifying gradients directly.
Representative methods include Noise~\cite{sok}, Pruning~\cite{sok}, Soteria~\cite{soteria}, Outpost~\cite{outpost}, and DGP~\cite{dgp}, which typically protect privacy by perturbing the magnitude of gradient parameters.

\section{Methodology}
\subsection{Defense Setup and Motivation}
\begin{figure}[t!]
    \centering
    \includegraphics[width=0.32\textwidth]{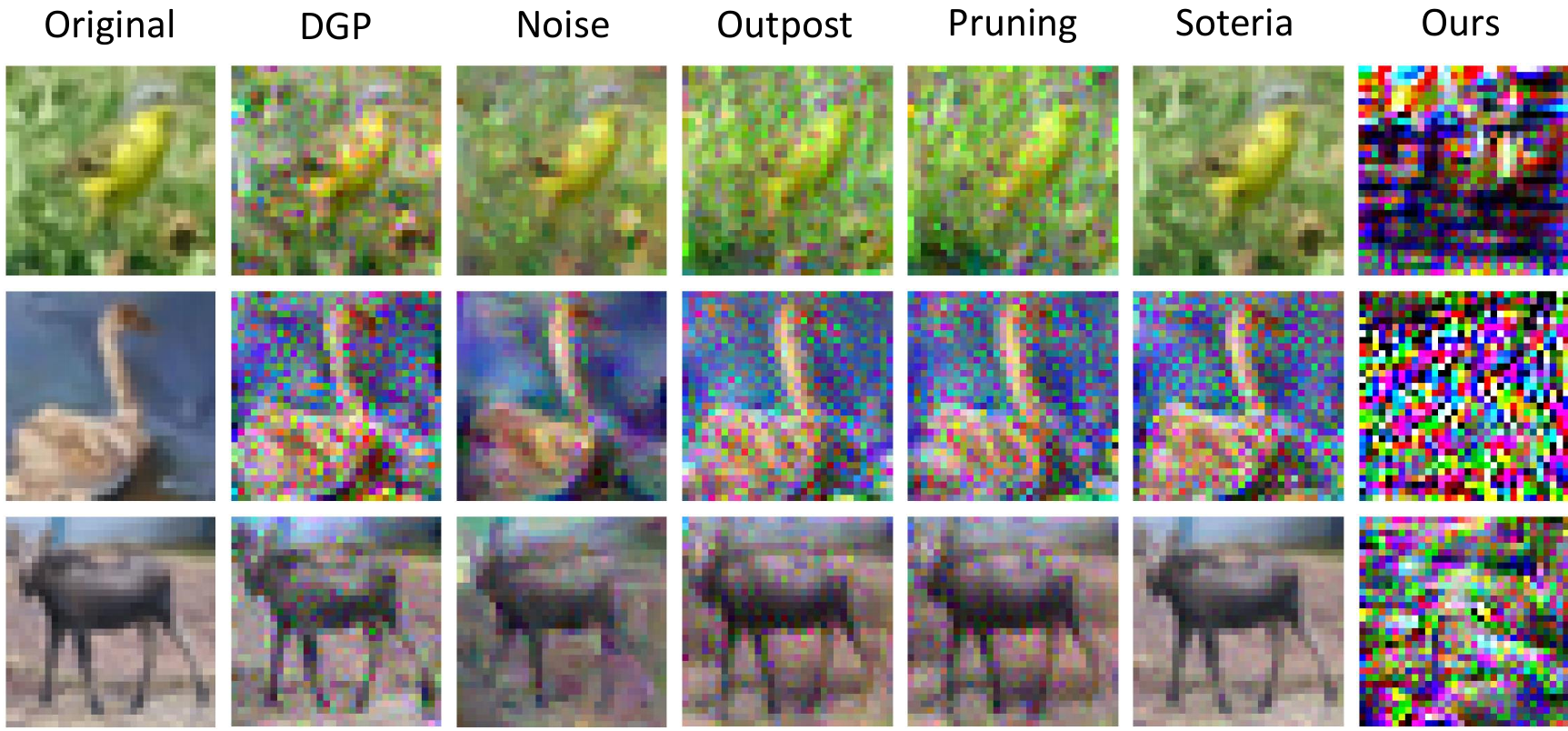}
        \centering
\caption{Visualization of defense results of various methods against UIA \protect\cite{recon1}, with CIFAR10 \protect\cite{cifar} and ResNet18 \protect\cite{resnet18}.}
  \label{fig:diff_model1}
  \vspace{-1.3em}
\end{figure}
\textbf{Defense Setup.} 
We adopt the threat model from~\cite{recon1}, where the attacker accesses both the original model \(\mathbf{\theta}_o\) and the unlearned model \(\mathbf{\theta}_u\), and performs UIA via DiffParm to recover forgotten data.
We assume the server acts as the defender, injecting perturbations into \(\mathbf{\theta}_u\) without altering the unlearning process. The defense ensures \textbf{privacy protection} by defending against adversarial unlearning inversion attacks, while achieving \textbf{utility preservation} by maintaining the accuracy and forgetting performance of the perturbed model \(\mathbf{\theta}_u'\).


\noindent \textbf{Motivation.}
Since prior work~\cite{recon1} has shown that the risk of unlearning inversion stems from the gradient information embedded in DiffParm, we explore the possibility of applying existing perturbation-based gradient inversion defenses to unlearning scenarios for privacy protection. We do not consider data mixing-based defenses, as they are not suitable for the unlearning scenarios.
As illustrated in Fig.~\ref{fig:diff_model1}, these methods perform poorly against unlearning inversion attacks. The underlying reason is that current defenses primarily focus on perturbing the magnitude of gradients and lack dedicated protection for direction-sensitive information in cosine space, which is precisely where unlearning inversion attacks reconstruct forgotten data. To address this limitation, we propose UnlearnShield, the first targeted defense specifically designed for unlearning inversion. UnlearnShield applies the optimized perturbations to the unlearned model to realize the defender’s objectives, incorporating dedicated designs for privacy protection and for utility preservation.

\subsection{Privacy Protection Design.}
\label{sec:privacy}
As shown in Sec.~\ref{related_work}, the attacker performs UIA by optimizing a synthetic sample such that its corresponding \({\Delta}^1\) minimizes the cosine distance to the true DiffParm \(\Delta\):
\begin{equation}
\min \; 1 - \frac{\langle \Delta^1, \Delta \rangle}{\|\Delta^1\|_2 \cdot \|\Delta\|_2},
\end{equation}
where \(\|\cdot\|_2\) denotes the \(\ell_2\)-norm.
This motivates our privacy module, which is based on the following intuition: given the true DiffParm \(\Delta\) and a perturbation \(\mathbf{\delta}\), we construct a perturbed  DiffParm \(\Delta^* = \Delta + \mathbf{\delta}\). A larger cosine distance (i.e., lower cosine similarity) between \(\Delta^*\) and \(\Delta\) misleads the attacker away from the correct direction, thereby enhancing privacy.
Therefore, we design the perturbation \(\mathbf{\delta}\) to minimize the cosine similarity between \(\Delta\) and \(\Delta^*\):
\begin{equation}
\min_{\mathbf{\delta}} \; \frac{\langle \Delta, \Delta^* \rangle}{\|\Delta\|_2 \cdot \| \Delta^*\|_2}, \quad \text{where } \Delta^* = \Delta + \mathbf{\delta}.
\end{equation}
Since cosine similarity ranges from \([-1, 1]\), we shift the target value to the positive range to avoid the vanishing gradient problem during optimization.
Therefore, we defend against UIA by utilizing the following privacy function:
\begin{equation}
\mathcal{L}_{\text{privacy}} = 1 + \frac{\langle \Delta, \Delta^* \rangle}{||\Delta||_2  ||\Delta^*||_2}, \quad \text{where } \Delta^* = \Delta + \mathbf{\delta}.
\label{equ:privacy}
\end{equation}

\subsection{Usability Design.}
\label{sec:usable}
The usability of the machine unlearning mechanism includes both model accuracy and the forgetting effect, and we have designed separate solutions for each.

\noindent\textbf{Accuracy Module.} 
We design our accuracy module that mitigates the negative impact of the perturbation on  accuracy by restricting the amplitude of \(\mathbf{\delta}\) during its initialization and optimization. Specifically, we apply the amplitude-based initialization model (AIM) derived from DiffParm to ensure that the perturbation \(\mathbf{\delta}\) remains within a reasonable range, as follows:
\begin{equation}
\label{equ:init}
\mathbf{\delta} \sim \mathcal{N}(0, \sigma^2), \quad \sigma = \frac{1}{n} \sum_{i=1}^n |\Delta_i|,
\end{equation}
where \( n \) is the number of elements in \( \Delta \).
During the optimization phase, we further constrain the magnitude of the perturbation to minimize its negative impact on the model's performance. The corresponding loss function is defined as:
\begin{equation}
\label{equ:acc}
\mathcal{L}_{acc}= \|\/\mathbf{\theta}_u^{'}-\mathbf{\theta}_u\|_2^2=\|\mathbf{\delta}\|_2^2.
\end{equation}
This function ensures that the amplitude of the perturbation remains controlled throughout the optimization process.

\noindent\textbf{{Forgetting Module.}}
We design a forgetting module to ensure that the perturbed unlearned model maintains the forgetting effect on the forgotten data.
Prior research~\cite{o1,o2} show that the forgetting effect of $\mathbf{x}$ can be reflected in the model’s output on $\mathbf{x}$: if the model aims to forget information related to $\mathbf{x}$, then its output for $\mathbf{x}$ should resemble the output of a model that has never seen $\mathbf{x}$, which we denote as ${\theta}_u^*$.
Since ${\theta}_u^*$ is generally unavailable in practice, as it would require retraining, we adopt the triangle inequality as an alternative strategy:
\begin{align}
\| f_{{\theta}_u^*}(\mathbf{x}) - f_{{\theta}_u'}(\mathbf{x}) \|_2
&\le
\| f_{{\theta}_u^*}(\mathbf{x}) - f_{{\theta}_u}(\mathbf{x}) \|_2 \notag \\
&\quad+
\| f_{{\theta}_u}(\mathbf{x}) - f_{{\theta}_u'}(\mathbf{x}) \|_2 .
\end{align}
Here, \(f_{{\theta}_u}(\cdot)\), \(f_{{\theta}_u'}(\cdot)\), 
and \(f_{{\theta}u^*}(\cdot)\) denote the output functions of \({\theta}_u\), \({\theta}_u'\), 
and \({\theta}_u^*\).
This inequality shows that minimizing the output difference between 
\(f_{{\theta}_u}(\mathbf{x})\) and \(f_{{\theta}_u'}(\mathbf{x})\) 
effectively preserves the forgetting effect by implicitly controlling deviation 
from the ideal model \(f_{{\theta}_ u^*}(\mathbf{x})\), without requiring access to 
\({\theta}_u^*\). Based on this insight, we define the following forgetting 
consistency loss:

\begin{equation}
\label{equ:forget}
\mathcal{L}_{\text{forget}}
=
\bigl\| f_{{\theta}_u}(\mathbf{x}) - f_{{\theta}_u'}(\mathbf{x}) \bigr\|_2.
\end{equation}
We use the output difference (OutDiff),
\(\mathrm{OutDiff}(\mathbf{x}) = \bigl\| f_{{\theta}_u}(\mathbf{x}) - f_{{\theta}_u'}(\mathbf{x}) \bigr\|_2\), as a metric to quantify the forgetting effect, the smaller the OutDiff, the better the forgetting effect.

\subsection{UnlearnShield Framework}
By combining the privacy and usability designs mentioned above, we propose {UnlearnShield}, which defends against unlearning inversion while ensuring effectiveness in unlearning scenarios. Specifically, we initialize the perturbation \( \mathbf{\delta} \) using formula~(\ref{equ:init}), and optimize using the following total loss:
\begin{equation}
\label{equ:all_loss}
\mathcal{L}_{\text{total}} = \mathcal{L}_{\text{privacy}} + \lambda_1 \mathcal{L}_{\text{acc}} + \lambda_2 \mathcal{L}_{\text{forget}},
\end{equation}
where \( \lambda_1 \) and \( \lambda_2 \) are hyperparameters. 

\section{Experiments}
\label{sec:experiments}
\subsection{Experimental Setup}
We conduct experiments in Python 3.8 with a single GeForce RTX 3090 GPU. Following~\cite{recon1}, we conduct evaluations on CIFAR10~\cite{cifar} and STL10~\cite{stl10}, using ResNet18~\cite{resnet18} as the default model.
Following~\cite{recon1}, we adopt the GA~\cite{GA1} as the default unlearning method. 
We consider single–data-point unlearning, which is most vulnerable to UIA. 
We include a no-defense baseline and compare against perturbation-based defenses: Soteria, OutPost, DGP, Pruning, and Noise.
For Noise, we add Gaussian noise with standard deviation \(0.1\); for Pruning, we prune \(90\%\) of gradient entries. Other defenses follow their original configurations.
For {UnlearnShield.}
We optimize with Adam using \(\lambda_1=0.5\), \(\lambda_2=0.5\), learning rate \(\eta=1\times10^{-5}\), and 10 iterations.
Following \cite{dgp}, we evaluate privacy using LPIPS and SSIM, where a higher LPIPS value and a lower SSIM value indicate better defense. 
We evaluate the model's accuracy on the test set. We use OutDiff to measure the forgetting effect. 
We randomly sample 100 images and report the mean of all metrics.

\begin{figure}[t]
  \centering
  \begin{subfigure}{0.31\textwidth}
    \centering
    \includegraphics[width=\linewidth]{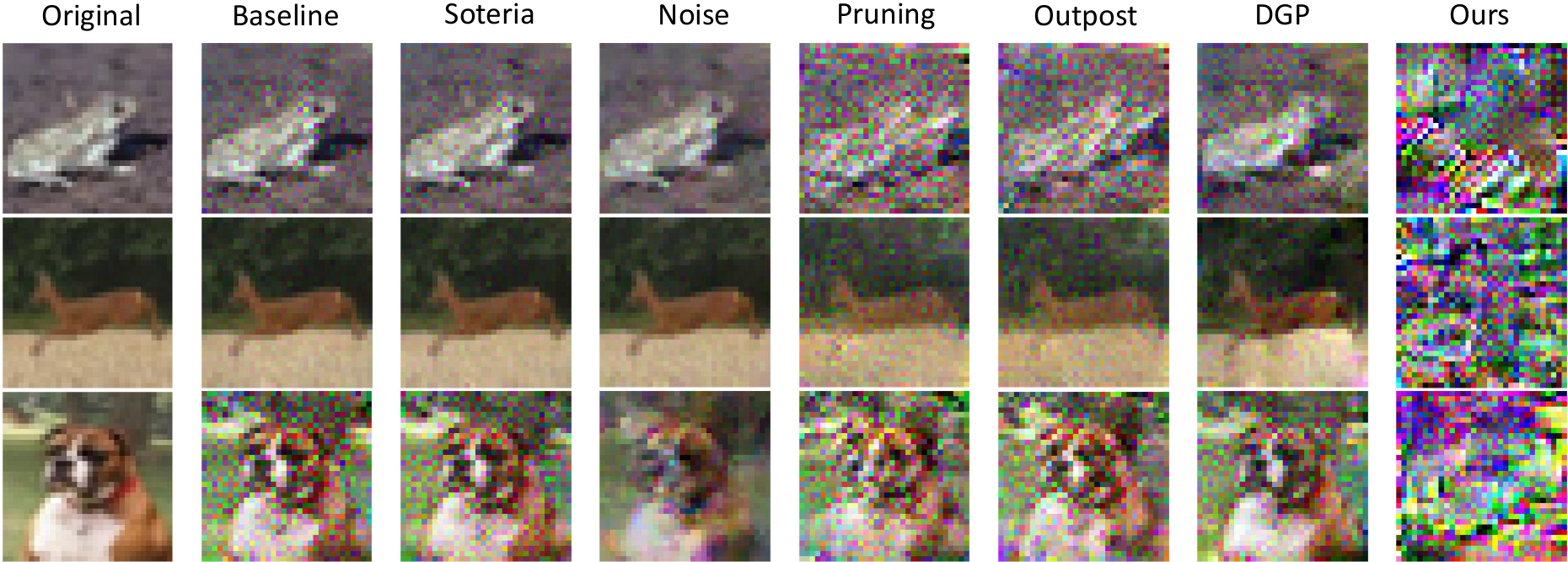}
    \subcaption{CIFAR10, ResNet18}\label{fig:privacy:c10}
  \end{subfigure}\hfill
  \begin{subfigure}{0.31\textwidth}
    \centering
    \includegraphics[width=\linewidth]{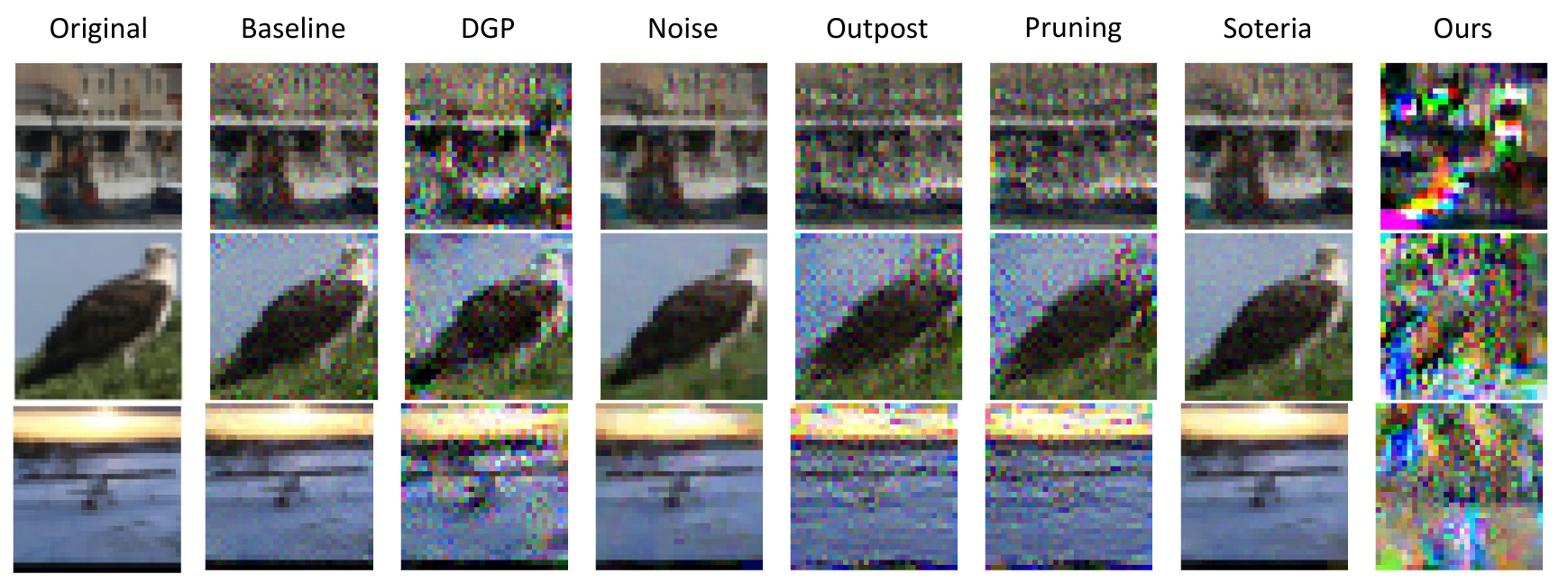}
    \subcaption{STL10, ResNet18}\label{fig:privacy:stl}
  \end{subfigure}
  \caption{Visualization results under UIA.}
  \label{fig:privacy}
  \vspace{-1.5em}
\end{figure}

\begin{figure*}[t]
  \centering
  \begin{subfigure}{0.23\textwidth}
    \centering
    \includegraphics[width=\linewidth]{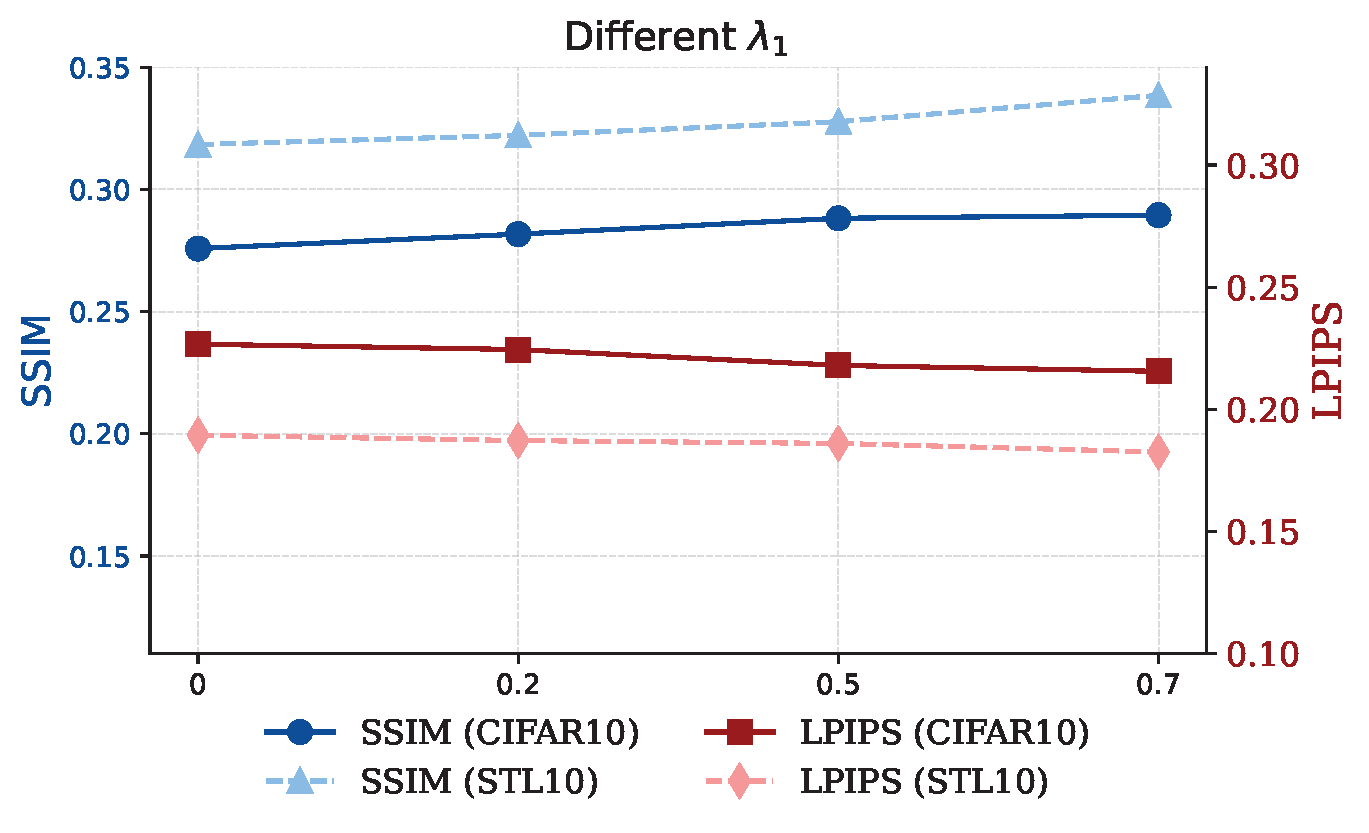}
    \subcaption{Privacy, $\lambda_1$}
  \end{subfigure}\hfill
  \begin{subfigure}{0.23\textwidth}
    \centering
    \includegraphics[width=\linewidth]{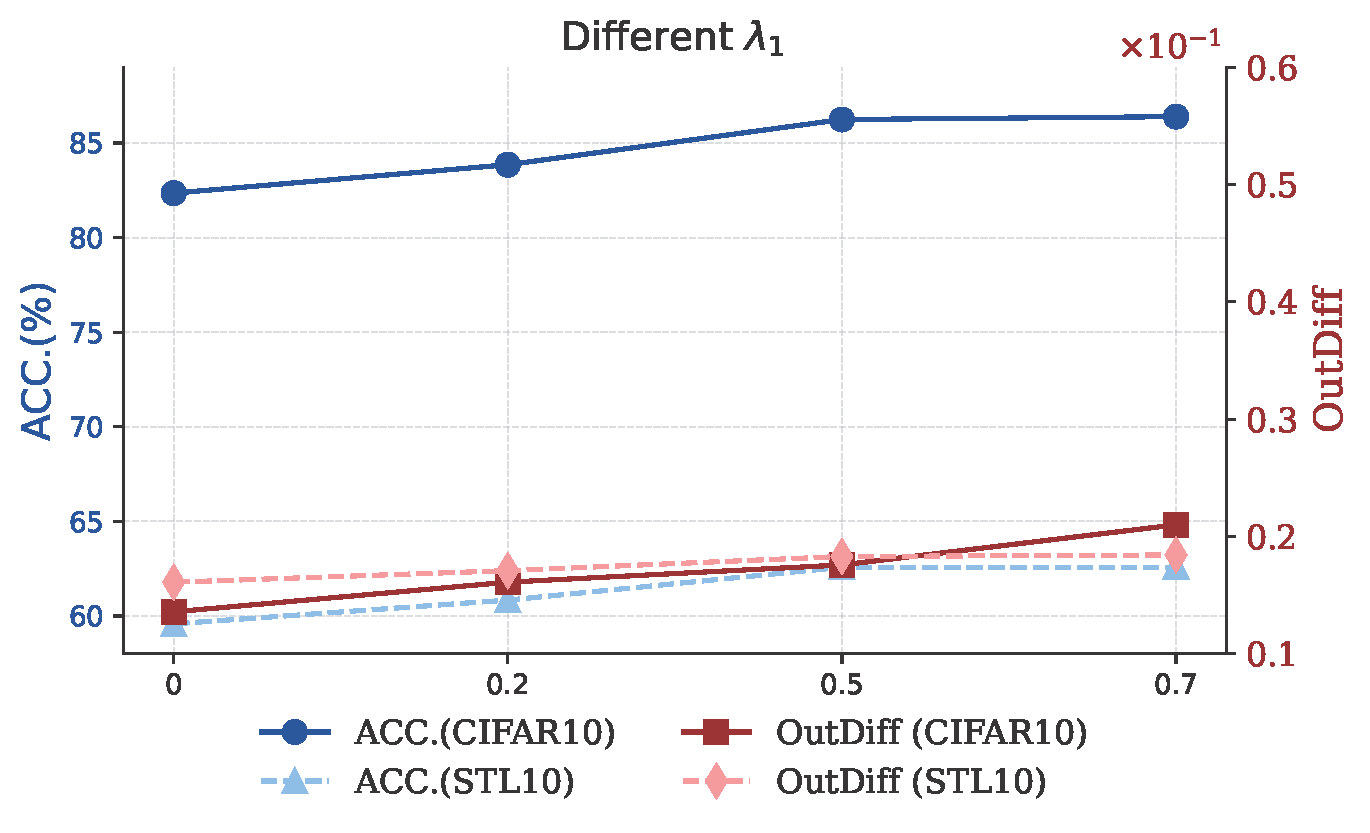}
    \subcaption{Usability, $\lambda_1$}
  \end{subfigure}\hfill
  \begin{subfigure}{0.23\textwidth}
    \centering
    \includegraphics[width=\linewidth]{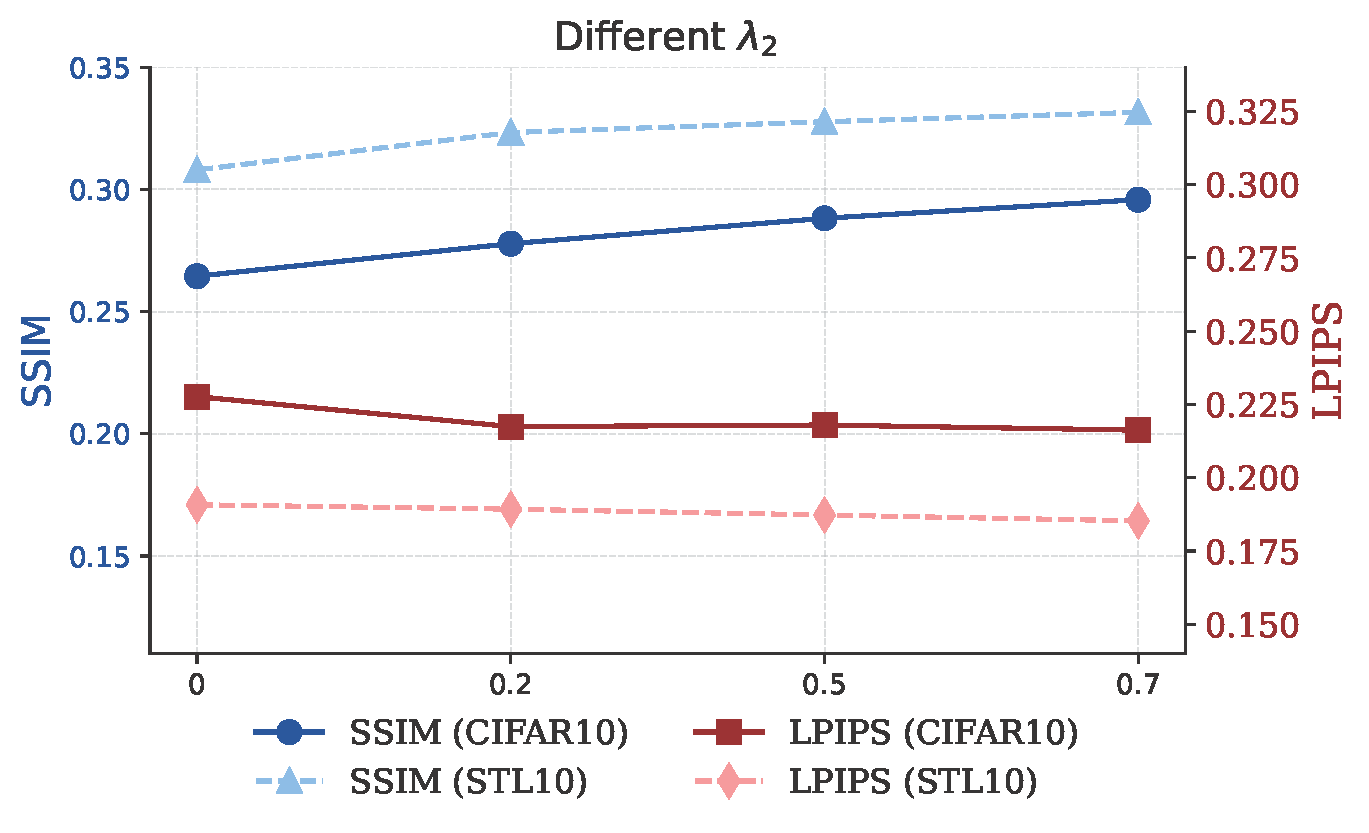}
    \subcaption{Privacy, $\lambda_2$}
  \end{subfigure}\hfill
  \begin{subfigure}{0.23\textwidth}
    \centering
    \includegraphics[width=\linewidth]{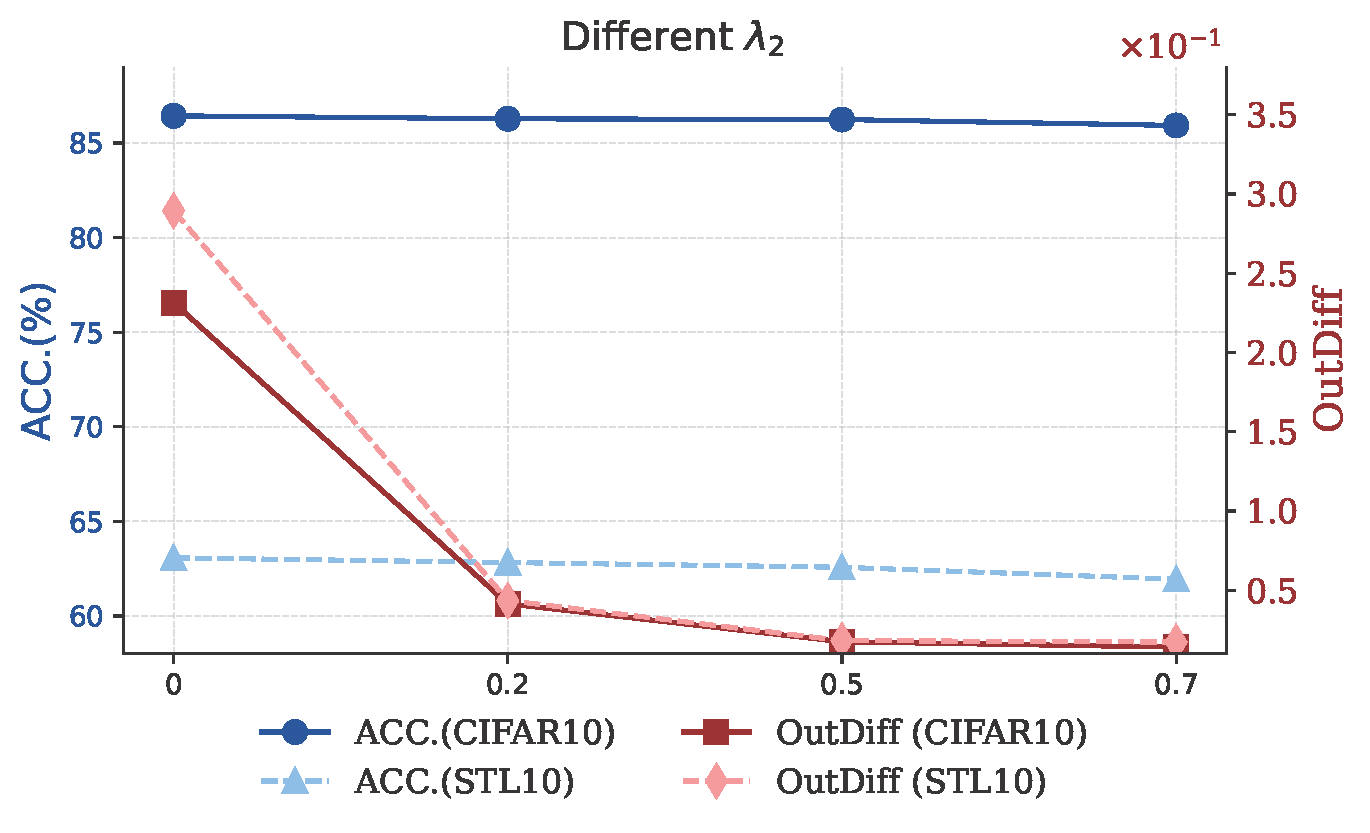}
    \subcaption{Usability, $\lambda_2$}
  \end{subfigure}
  \caption{Evaluation of the impact of different $\lambda_1$ and $\lambda_2$ on privacy and usability. Privacy is measured by SSIM~($\uparrow$) and LPIPS~($\downarrow$), while usability is measured by ACC~($\uparrow$) and OutDiff~($\downarrow$).}
  \label{fig:ab}
  \vspace{-0.5em}
\end{figure*}

\begin{table*}[htbp]
  \centering
  \caption{Evaluate the performance of different defenses, with the best results highlighted in bold.}
   \scalebox{0.63}{
     \begin{tabular}{llllllllllllllll}
    \toprule
    \multirow{2}[4]{*}{} & \multirow{2}[4]{*}{Metric} & \multicolumn{7}{c}{CIFAR10}                    & \multicolumn{7}{c}{STL10} \\
\cmidrule(lr){3-9}   \cmidrule(lr){10-16}       &      & Baseline & DGP  & Noise & Outpost & Pruning & Soteria & Ours & Baseline & DGP  & Noise & Outpost & Pruning & Soteria & Ours \\
   \cmidrule(lr){1-9}   \cmidrule(lr){10-16} 
    \multirow{2}[2]{*}{Privacy} & SSIM & 0.7530  & 0.6419  & 0.4736  & 0.5746  & 0.5613  & 0.7516  & \textbf{0.2882 } & 0.7261  & 0.5092  & 0.6117  & 0.5156  & 0.5075  & 0.6901  & \textbf{0.3277 } \\
         & LPIPS & 0.0135  & 0.0286  & 0.0988  & 0.0341  & 0.0374  & 0.0185  & \textbf{0.2180 } & 0.0143  & 0.0637  & 0.0673  & 0.0581  & 0.0593  & 0.0241  & \textbf{0.1873 } \\
     \cmidrule(lr){1-9}   \cmidrule(lr){10-16} 
    \multirow{2}[2]{*}{Usability} & Acc.(\%) & \textbf{86.734} & 85.411  & 10.000  & 86.012  & 86.212  & 86.435  & 86.238  & \textbf{62.655} & 61.575  & 9.166  & 61.760  & 62.520  & 62.579  & 62.570  \\
         & OutDiff & \textbf{0.0000 } & 0.0257  & 1.7159  & 0.0164  & 0.0114  & 0.0070  & 0.0176  & \textbf{0.0000} & 0.0226  & 1.7346  & 0.0187  & 0.0107  & 0.0009  & 0.0183  \\
    \bottomrule
    \end{tabular}
    }
  \label{tab:compare}%
  \vspace{-1em}
\end{table*}%

\subsection{Comparative Experiments}
Tab.~\ref{tab:compare} and Fig.~\ref{fig:privacy} demonstrate that, compared to other methods, only our method provides effective defense against unlearning inversion attacks. Furthermore, Tab.~\ref{tab:compare} shows that our method, along with Soteria, DGP, pruning, and Outpost, has minimal impact on the model's accuracy and forgetting performance, whereas Noise leads to significant degradation.
\subsection{Ablation Experiments}
\textbf{Ablation of \( \mathcal{L}_{\text{acc}} \) and  \( \mathcal{L}_{\text{forget}} \).}
As shown in Fig.~\ref{fig:ab}, increasing $\lambda_1$ improves accuracy but slightly degrades privacy and forgetting, while increasing $\lambda_2$ enhances forgetting but slightly reduces accuracy and privacy. Notably, our method achieves a favorable balance under most settings.
%

\noindent \textbf{Ablation of AIM.}
We perform this ablation by replacing AIM with Gaussian noise of standard deviation $\sigma = 0.01$, a common initialization setting~\cite{dgp}, denoted as Ours (w/o AIM). As shown in Tab.~\ref{tab:ab_aim}, AIM effectively preserves usability.

\begin{table}[t!]
  \centering
 \caption{Ablation evaluation for AIM module.}
   \scalebox{0.5}{
    \begin{tabular}{lrrrrrrrr}
    \toprule
    \multicolumn{1}{c}{\multirow{2}[4]{*}{Method}} & \multicolumn{4}{c}{CIFAR10} & \multicolumn{4}{c}{STL10} \\
\cmidrule(lr){2-5}  \cmidrule(lr){6-9}        & \multicolumn{1}{l}{SSIM } & \multicolumn{1}{l}{LPIPS} & \multicolumn{1}{l}{OutDiff} & \multicolumn{1}{l}{Acc. (\%)} & \multicolumn{1}{l}{SSIM } & \multicolumn{1}{l}{LPIPS} & \multicolumn{1}{l}{OutDiff} & \multicolumn{1}{l}{Acc. (\%)} \\
\cmidrule(lr){1-5}  \cmidrule(lr){6-9}   
    Ours & 0.2882  & 0.2180  & 0.0176  & 86.238  & 0.3277  & 0.1873  & 0.0183  & 62.570  \\
    Ours(w/o AIM) & 0.2798  & 0.2275  & 1.4209  & 21.411  & 0.2956  & 0.2041  & 1.4423  & 21.352  \\
    \bottomrule
    \end{tabular}%
    }
 \vspace{-0.5em}
  \label{tab:ab_aim}%
\end{table}%

\begin{table}[t!]
  \centering
  \caption{Evaluation under different unlearning methods.}
  \scalebox{0.5}{
    \begin{tabular}{lllllllll}
    \toprule
    \multirow{2}[4]{*}{Method} & \multicolumn{4}{c}{CIFAR10}   & \multicolumn{4}{c}{STL10} \\
\cmidrule(lr){2-5}  \cmidrule(lr){6-9}           & SSIM  & LPIPS & Acc.(\%) & OutDiff & SSIM  & LPIPS & Acc.(\%) & OutDiff \\
    \cmidrule(lr){1-5}  \cmidrule(lr){6-9}   
    Sparse & 0.2544  & 0.2700  & 85.880  & 0.0104  & 0.2575  & 0.2336  & 62.513  & 0.0188  \\
    Salun  & 0.3022  & 0.1565  & 86.730  & 0.0130  & 0.2649  & 0.2116  & 63.688  & 0.0173  \\
    \bottomrule
    \end{tabular}%
  \label{tab:diff_unlearn}%
  }
   \vspace{-1em}
\end{table}%

\subsection{Further discussion}
\label{sec:disc}
\noindent\textbf{Different unlearning methods.} 
Tab.~\ref{tab:diff_unlearn} shows that UnlearnShield also performs well under Salun~\cite{Salun} and Sparse~\cite{sparse}.
\textbf{More models.}
Tab.~\ref{tab:eva_cnn} presents the evaluation of our method using the ConvNet from \cite{IG} on CIFAR10 and STL10. The results show that our method still provides effective defense.
\begin{table}[t!]
  \centering
       \caption{Evaluation with ConvNet on different datasets.}
  \label{tab:eva_cnn}%
     \scalebox{0.5}{
   
    \begin{tabular}{cccrrccrr}
    \toprule
    \multirow{2}[4]{*}{Method} & \multicolumn{4}{c}{CIFAR10} & \multicolumn{4}{c}{STL10} \\
\cmidrule(lr){2-5}  \cmidrule(lr){6-9}          & SSIM & LPIPS & \multicolumn{1}{l}{Acc. (\%)} & \multicolumn{1}{l}{OutDiff} & SSIM & LPIPS & \multicolumn{1}{l}{Acc. (\%)} & \multicolumn{1}{l}{OutDiff} \\
 \cmidrule(lr){1-5}  \cmidrule(lr){6-9}   
    Baseline & 0.7295  & 0.0134  & 92.411  & 0.0000  & 0.7700  & 0.0087  & 63.904  & 0.0000  \\
    Ours & 0.2834  & 0.2394  & 92.126  & 0.0117  & 0.3222  & 0.2026  & 63.698  & 0.0182  \\
    \bottomrule
    \end{tabular}%
    }
    \vspace{-0.5em}
\end{table}%

\noindent\textbf{Membership inference.}
We further evaluate the forgetting effect using the MIA from \cite{Salun}, with AUC as the metric. On CIFAR10, the baseline achieves \textbf{0.1214} while our method achieves \textbf{0.1226}; on STL10, the baseline is \textbf{0.1597} and ours is \textbf{0.1635}. The close results indicate that our method preserves the forgetting effect as well as the baseline.
\begin{figure}[t!]
    \centering
    \includegraphics[width=0.32\textwidth]{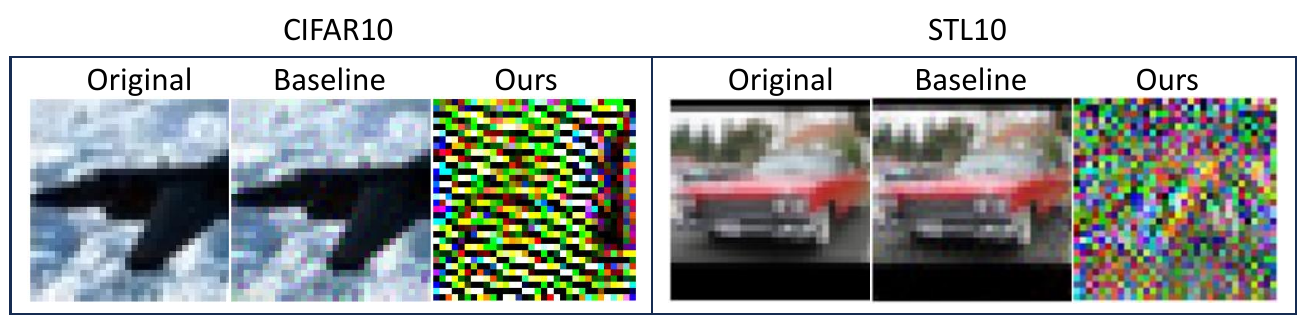}
        \centering
\caption{Visualization results under adaptive UIA.}
  \label{fig:ada}
  \vspace{-1.3em}
\end{figure}

\noindent\textbf{Adaptive UIA.}
We assume an adaptive attacker who is aware of our defense and optimizes virtual data to match the perturbed DiffParm. Despite this, UnlearnShield remains effective: on CIFAR10, SSIM and LPIPS are \textbf{0.3074} and \textbf{0.1887}; on STL10, \textbf{0.3014} and \textbf{0.1631}, respectively. These results and Fig.~\ref{fig:ada}, confirm its robustness against adaptive UIA.

\noindent\textbf{Computation cost.} Our method incurs an average comp. cost of 0.3898s on CIFAR10 and 0.4253s on STL10, which is lightweight and acceptable given its performance.

\section{Limitation and Conclusion}
\label{sec:conclusion}
This paper introduces UnlearnShield, the first dedicated defense against unlearning inversion risks. As a post-processing method, it operates without modifying the unlearning process. Experimental results show that UnlearnShield effectively disrupts data reconstruction while preserving model accuracy and forgetting effect, striking a favorable balance between privacy protection and utility preservation.
The limitation of this work is its focus on the computer vision domain. Future research will explore broader machine unlearning scenarios, aiming for more comprehensive privacy protection across diverse data types.
\vfill\pagebreak
\bibliographystyle{IEEEbib}
\bibliography{strings,refs}

\end{document}